# REFLECTIONLESS FILTER STRUCTURES

*Matthew A. Morgan and Tod A. Boyd*

**Abstract**

This paper expands on the previously described reflectionless filters – that is, filters having, in principle, identically-zero reflection coefficient at all frequencies – by introducing a wide variety of new reflectionless structures that were not part of the original publication. In addition to extending the lumped-element derivation to include transmission line filters, this is achieved by the introduction of a novel method wherein the left- and right-hand side stop-band terminations are coupled to each other using a two-port sub-network. Specific examples of the sub-network are given which increase the stop-band attenuation per filter cell and steepen the cutoff response without disrupting the reflectionless property or increasing the insertion-loss in the pass-band. It is noteworthy that a common feature of all the structures derived by these methods is that most, if not all, of the reactive elements are of equal normalized value. This greatly simplifies the tuning requirements, and has facilitated their implementation as monolithic microwave integrated circuits (MMICs). A number of examples of MMIC reflectionless filters constructed in this way are presented and their results compared to the theory.

## I. INTRODUCTION

The recently proposed reflectionless filter topology [1]-[4] provides a third-order Inverse-Chebyshev response while in principle having identically-zero reflection coefficient at all frequencies, including pass-band, stop-band, and transition-band. This can be beneficial in microwave system design where poor out-of-band termination may cause unwanted standing waves, instability, and issues with dynamic range. These structures have already been applied to broadband microwave downconverters where they stabilize the conversion loss in the presence of uncontrolled high-frequency terminations, and in active multiplier chains where they absorb intermediate harmonics to flatten out the frequency response. There has also been interest in integrating them closely with mixer ICs, and with Analog-to-Digital Converters (ADCs) in order to improve spurious free dynamic range via the absorption of out-of-band digital noise.

The original reflectionless filter structures [1], which will be reviewed in Section II, are restricted to lumped-element realizations. Attempts have been made to convert these to transmission line implementations, better suited to high frequency designs, but complete conversion of all reactive elements to transmission lines has proven unusually challenging [4]. This has now been achieved, and the solution is presented in Section III.

Further, the frequency response of the original lumped element structures does have some disadvantages. Constraints imposed by the reflectionless condition fix the element values and consequently the pole/zero-frequencies, thus limiting the slope of the cutoff response and the peak stop-band attenuation to 14.47 dB per cascaded cell. A method for overcoming these deficiencies is presented in Sections IV-VIII, wherein the matched terminations which absorb stop-band energy from the left and right sides, respectively, are replaced with a two-port, matched sub-network coupling the input side to the output side. Since this network is only active in the stop- and transition-bands, no additional loss is incurred in the pass-band, and since it is matched, the reflectionless property remains intact. However, a judicious choice of the sub-network's amplitude and phase characteristic may be used to increase the transition slope and to reduce the 14.47 dB leakage through the filter proper. Finally, it is important to note that while many examples in this paper will for convenience be limited to low-pass filters, the same general results will apply to high-pass, band-pass, band-stop, and multi-band implementations as well, the structures for which may be readily obtained by well known lumped-element transformations.

## II. Original Reflectionless Filter Derivation

As described in [1], the original reflectionless filter was derived from first principles under the assumption that the structure would be symmetric. Even-/Odd-Mode Analysis provides the framework. It is known that the behavior of such a symmetric two-port network can be fully described by its response to two standard excitations: an even-mode, wherein the two ports are driven equal in amplitude and in-phase, and an odd-mode, wherein the two ports are driven with equal amplitude, but out-of-phase. Symmetry allows the network to be divided in half, assigning an open-circuit to the nodes on the symmetry plane in the case of the even-mode, or to ground (sometimes called a virtual short) in the case of the odd-mode. The reflection coefficient of the full two-port will then be the superposition of the reflection coefficients of the even- and odd-mode equivalent circuits, whereas the transmission coefficient will be proportional to the difference. That is,

$$s_{11} = \tfrac{1}{2}(\Gamma_{even} + \Gamma_{odd}) \qquad (1a)$$
$$s_{21} = \tfrac{1}{2}(\Gamma_{even} - \Gamma_{odd}). \qquad (1b)$$

Thus, for the circuit to be reflectionless ($s_{11}$=0), we must ensure that the even- and odd-mode reflection coefficients are equal in amplitude and opposite in sign. We therefore begin by drawing the even-mode equivalent circuit as a terminated high-pass filter, and the odd-mode equivalent circuit as its dual, as shown in Fig. 1a. The concept of duality traces back to the symmetry of the electric and magnetic field terms in Maxwell's Equations, and the circuit dual is constructed in this context by exchanging inductors with capacitors, and series elements with shunt elements, etc. This will ensure that the even- and odd-modes will have equal amplitude reflection coefficients with the opposite sign.

Having provided for the identically-zero reflection coefficient and the desired transmission response, it remains only to redraw the equivalent half-circuits with the required symmetry to make (1) valid. The first step is to exchange the positions of the series resistor and capacitor in the even-mode equivalent circuit, and change the grounding of some elements in the odd-mode equivalent circuit from absolute to virtual, as shown in Fig. 1b. Step two, shown in Fig. 1c, involves adding open-circuited elements to the even-mode side (keeping in mind that the symmetry plane is an open-circuit for the even-mode) and a capacitor shorted at both ends on the odd-mode side (for which the symmetry plane is a short-circuit). The completed filter is thus shown in Fig. 1d.

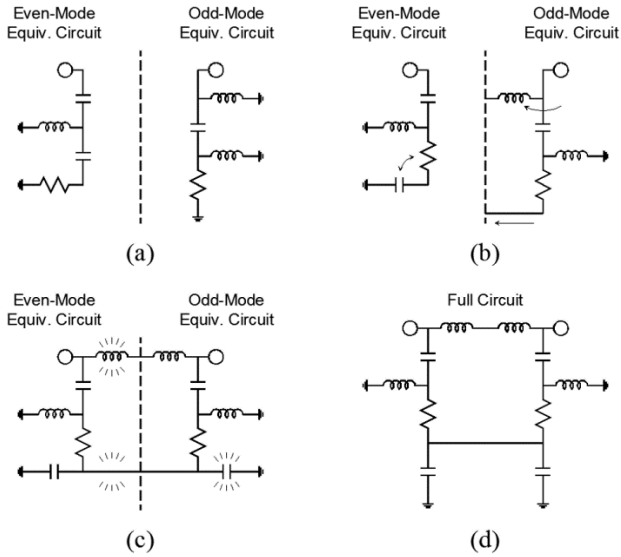

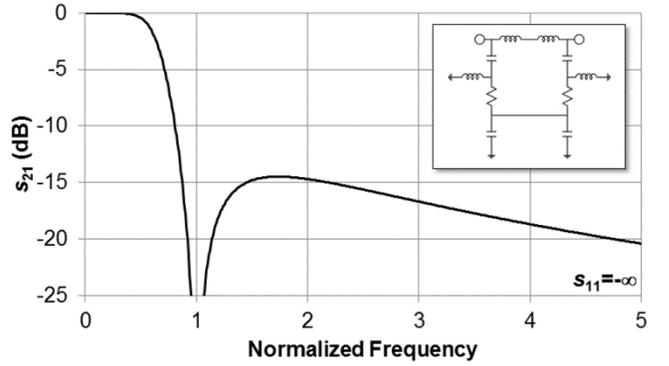

Fig. 2. Simulated transmission coefficient of the reflectionless filter in Fig. 1d. The reflection coefficient in decibels is infinite.

Fig. 1. Derivation of the original first-order (low-pass) reflectionless filter topology. a) Dual, terminated high-pass filters as even- and odd-mode equivalent circuits. b) Swapping series *R* and *C*, and changing ground nodes from absolute to virtual. c) Addition of open-ended and short-circuited elements. d) Final reflectionless filter topology. All inductors $L=Z_0/\omega_z$, capacitors $C=Y_0/\omega_z$, and resistors $R=Z_0$, where $\omega_z$ is the frequency of the attenuation zero.

Symmetry and duality constrain the values that the circuit elements may have. For example, the first inductor in the odd-mode equivalent circuit in Fig. 1a must be the dual of the first capacitor in the even-mode equivalent circuit (that is, $L/C = Z_0^2$). By symmetry, however, that capacitor must be equal to its counterpart on the odd-mode side. This in turn is the dual of the next element (the inductor) on the even-mode side, and so on. No matter what the order of the prototype ladder-network, the end result is a chain of equations whereby $L_1/Z_0=C_1/Y_0=L_2/Z_0$, etc., leading eventually to the conclusion that all reactive elements must have the same values. If we normalize these such that $\omega_z^2 LC=1$, we may conclude that all inductors $L=Z_0/\omega_z$, and all capacitors $C=Y_0/\omega_z$, where $\omega_z$ is the frequency of the attenuation zero. The two resistors must be identical and their own dual, so $R=Z_0$. For simplicity, these values will be assumed for all circuit schematics in this paper unless otherwise noted. The simulated frequency response of this circuit is shown in Fig. 2.

### III. Transmission line Reflectionless Filter

The foregoing derivation for basic lumped-element filters was described in [1]. No explanation was given, however, for how such filters could be converted to transmission line form. This proved over time to be unexpectedly challenging, as the series stubs (which are generally considered undesirable for most printed-circuit implementations) became "trapped" between circuit nodes in such a way that Kuroda's Identities did not apply [4]. Only recently has a complete solution to this problem been found, and the resultant derivation is summarized in Fig. 3.

As before, the derivation begins with an even-mode equivalent filter drawn to establish the frequency response, and the odd-mode equivalent circuit which is its dual. These are shown in Fig. 3a. Richard's Transformations [5] are applied in Fig. 3b to convert the reactive elements to transmission line stubs. At this stage, the resistive terminations are also offset by a quarter-wavelength matched transmission line. (All transmission line sections and stubs here and elsewhere in this paper will be assumed to be a quarter-wavelength long at the designed center frequency.) Kuroda's Identity is applied in Fig. 3c to convert series-connected short-circuited stubs to shunt-connected open-circuited stubs, and vice-versa. In Fig. 3d, the series stub and termination resistor on the even-mode side are swapped (effectively turning this series stub into a shunt stub) and the grounding of some elements on the odd-mode side are converted from absolute to virtual (in this case, turning a shunt stub into a series stub). Next, some open-circuited elements are added to the even-mode side in Fig. 3e, while a stub that is short-circuited at both ends is added to the odd-mode side.

To eliminate the two "captive" series stubs at the top of the circuit requires a little-known transmission line identity illustrated in Fig. 4. No prior reference for this identity has been found, but it will be shown in the Appendix that the three-port network on the left-hand side has identical network parameters as the coupled-line network on the right-hand side, where all lines have the same electrical length, with the following element values,

$$Z_{odd} = Z_1\left(1 - \tfrac{1}{n}\right) \quad (2a)$$

$$Z_{even} = Z_1\left(1 + \tfrac{1}{n}\right) \quad (2b)$$

$$n = \sqrt{\tfrac{Z_2}{Z_1} + 1} \ . \quad (2c)$$

The final transmission line reflectionless filter after application of this identity is shown in Fig. 3f. Note that the back-to-back transformers one would need to introduce into the network to strictly apply this identity cancel each other out. As before, the symmetry and duality required for reflectionless operation constrains the element values. For a given coupling coefficient, $\rho$, we have:

$$Z_{even} = Z_0 \sqrt{\rho} \tag{3a}$$

$$Z_{odd} = Z_0 / \sqrt{\rho} \tag{3b}$$

$$Z_1 = 2Z_0 \sqrt{\rho} \frac{\rho+1}{(\rho-1)^2} \tag{3c}$$

$$Z_2 = \frac{Z_0^2}{Z_1}. \tag{3d}$$

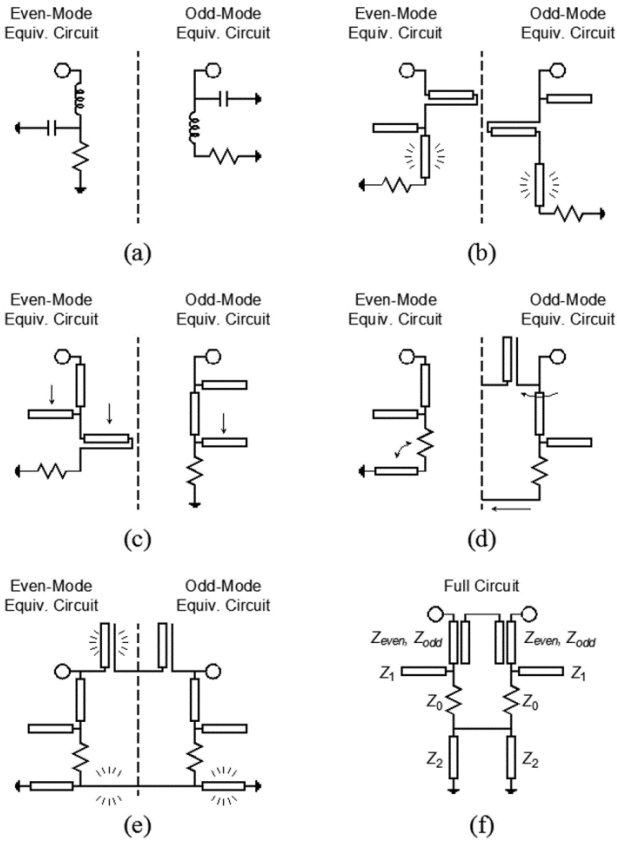

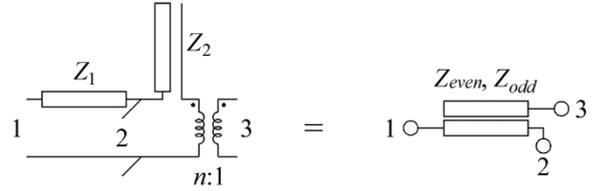

Fig. 4. Identity used in the derivation of the transmission line reflectionless filter. All transmission lines have the same electrical length.

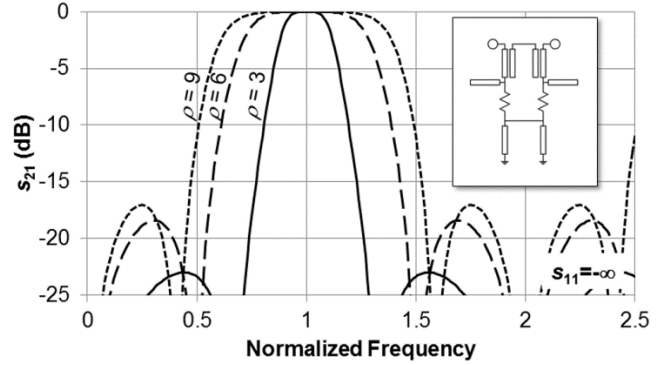

Fig. 5. Simulated transmission coefficient of the transmission line reflectionless filter of Fig. 3f. The reflection coefficient in decibels is infinite. Element values are given in the text.

Fig. 3. Derivation of the transmission line first-order band-pass reflectionless filter topology. a) Dual, terminated low-pass filters as even- and odd-mode equivalent circuits. b) Application of Richard's Transformations converting reactive elements to transmission line stubs, and adding a matched quarter-wave transmission line to the resistive terminations. c) Application of Kuroda's Identity. d) Swapping series short-circuited stub with termination resistor, and changing ground connections from absolute to virtual. e) Addition of open-ended and short-circuited elements. f) Final filter topology after application of the coupled-line identity in Fig. 4.

The simulated frequency response for several coupling coefficients is shown in Fig. 5. As with all commensurate line networks, the ideal frequency response is periodic, leading to a characteristic which may be described as band-pass, despite starting with a low-pass lumped-element prototype. A comparable solution for distributed band-stop filters (starting from a high-pass lumped-element prototype) has not been found at this time.

## IV. The Reflectionless Filter as a Diplexer

Returning to the lumped-element filter in Fig. 1d, it can be shown that all stop- and transition-band signal power absorbed by the circuit from the left input port is dissipated in the resistor on the left side, and all signal power absorbed from the right port is dissipated in the resistor on the right side. That is, the right-side resistor is isolated from the left input, and vice-versa. With this in mind, it is useful to consider the filter as a dual-directional diplexer wherein the stop-band energy is routed to internal ports which are nominally terminated by these resistors. The low-pass filter circuit is redrawn in Fig. 6 to highlight this conceptualization.

Intuitively, one would think that the maximum stop-band attenuation is achieved when the stop-band ports are simply terminated, as originally derived. However, some amount of stop-band energy is still allowed to leak through the other leg of the diplexer, in this case the inductors across the top. If a comparable signal amplitude were coupled through to the output via an internal sub-network, but out-of-phase with the primary leakage path, then an overall increase in stop-band attenuation could be achieved. Moreover, if the alternate signal path could be tailored to add constructively with the leakage at the start of the transition-band, but destructively at the end of it, an overall improvement in the steepness of the transition could be realized.

Before attempting to design a sub-network, it is useful to write out the scattering-parameters of the four-port diplexer/filter in mathematical form. To do so, we employ Even-/Odd-mode Analysis as described previously. The consequent odd-mode equivalent circuit in this relatively simple example is a pi-network of shunt inductors and a series capacitor, and that of the even-mode is a tee-network of the same (where in the latter case the positions of port 3 and the final series capacitor, which are in series, have been swapped). Using the Laplacian frequency parameter $s=\sigma+j\omega$, normalized such that the filter attenuation zero $\omega_z=1$, the normalized even-mode input impedance looking into port 1 is given by

$$z_e = \tfrac{1}{s} + \left(\tfrac{1}{s} + \left(\tfrac{1}{s}+1\right)^{-1}\right)^{-1} = \tfrac{s^3+2s^2+s+1}{s^3+s^2+s} \qquad (4)$$

and thus the transfer function of the low-pass filter is given by

$$H(s) = s_{21} = \tfrac{z_e-1}{z_e+1} = \tfrac{s^2+1}{(2s^2+s+1)(s+1)}. \qquad (5)$$

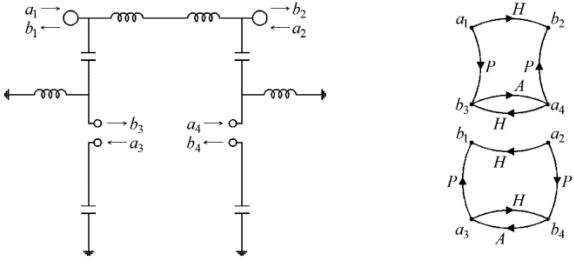

Fig. 6. First-order reflectionless filter topology, redrawn to emphasize the interpretation as a dual-directional diplexer having internal ports 3 and 4, formerly terminated with resistors. In the signal-flow graph on the right, $A$ is the insertion loss of the matched sub-network connected between ports 3 and 4, if present.

The remaining s-parameters may be derived using similar techniques,

$$S(s) = \begin{bmatrix} 0 & H(s) & P(s) & 0 \\ H(s) & 0 & 0 & P(s) \\ P(s) & 0 & 0 & H(s) \\ 0 & P(s) & H(s) & 0 \end{bmatrix} \qquad (6a)$$

$$= \tfrac{1}{(2s^2+s+1)(s+1)} \begin{bmatrix} 0 & s^2+1 & 2s^3 & 0 \\ s^2+1 & 0 & 0 & 2s^3 \\ 2s^3 & 0 & 0 & s^2+1 \\ 0 & 2s^3 & s^2+1 & 0 \end{bmatrix}. \qquad (6b)$$

The signal-flow graph for the 4-port circuit is shown on the right of Fig. 6, where it is assumed that a matched, sub-network has been connected bridging ports 3 and 4 with a transfer function of $A(s)$. Note that two independent graphs are obtained for the forward and reverse transmission-coefficients, respectively. The net effective transfer function for the entire circuit with embedded sub-network is therefore given by

$$H'(s) = H(s) + \tfrac{A(s)P^2(s)}{1-A(s)H(s)}. \qquad (7)$$

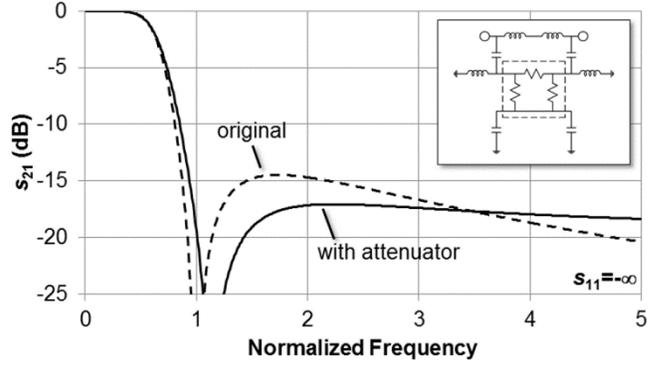

Fig. 7. Simulated performance of a reflectionless low-pass filter employing a simple attenuator as the sub-network (inset). The attenuation selected for maximum peak stop-band rejection is 19.43 dB.

This equation may be used to guide the development of sub-networks to achieve a desired characteristic.

## V. Some Useful Sub-Networks

Perhaps the most basic implementation of this idea is to use a simple resistive attenuator as the sub-network, as shown in Fig. 7. While it can improve the peak rejection by a couple of dB, the phase does not match the nominal stop-band leakage well enough for better cancellation. For the addition of a single resistor to the original circuit, it is of interest primarily for its simplicity and instructional value. More sophisticated examples will follow offering more significant improvements.

In principle, any circuit can be used as a sub-network to modify the filter response. As long as this sub-network itself is impedance-matched, so will be the overall circuit. An interesting choice is to use another reflectionless filter in this role. Serendipitously, if the inner filter is tuned to the same frequency as the outer, this has the effect of adding constructively to the leakage in the first part of the transition-band, but less so in the last part. The end result is a moderately steeper filter cutoff, as shown in Fig. 8.

One may continue embedding reflectionless filters inside one another in this fashion indefinitely, resulting (typically) in steeper cutoffs but higher stop-band peaks as well. They may also be cascaded with additional filter cells tuned to any frequency, with embedded attenuators or with other matched circuits, constructing ever more elaborate sub-networks in order to tailor the response for a particular application. The number of possible variations is innumerable, but one that stands out is a third-order nested reflectionless filter (that is, a filter embedding a filter embedding a filter) as shown in Fig. 9. In this example, the nested filters are all tuned to the same frequency, so every inductor has the same value, as does every capacitor and every resistor, greatly simplifying the tuning of these filters compared to conventional topologies with similar complexity. The innermost filter contains only termination resistors in its sub-network. Note that the transition is quite steep, and although the peak stop-band rejection is fairly poor at ~10 dB, it is much nearer to 20 dB for the first octave beyond cutoff, making it well-suited as an anti-aliasing filter wherein potential spurious signals in the lowest order Nyquist zones are typically the most important. As always, the circuit remains reflectionless, and therefore multiple copies may be cascaded for any total stop-band attenuation that one desires. One may also include a simple (first-order) reflectionless filter with a higher cutoff-frequency in the cascade to improve the rejection in the extended stop-band beyond the first Nyquist zone.

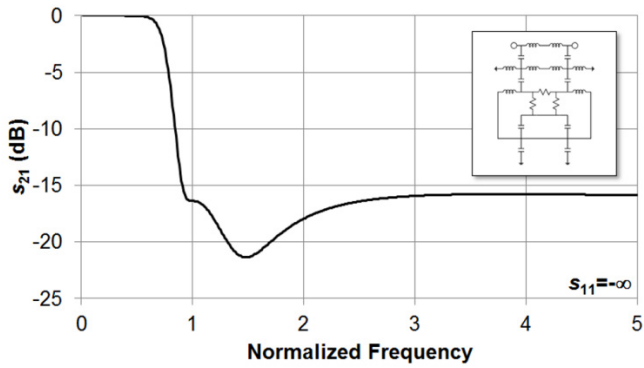

Fig. 8. Simulated performance of a reflectionless low-pass filter embedding another reflectionless filter as its sub-network (inset). The innermost filter in turn embeds a pi-attenuator. Results are shown with the attenuator designed for 16.39 dB, found to be optimal in this case for peak stop-band rejection.

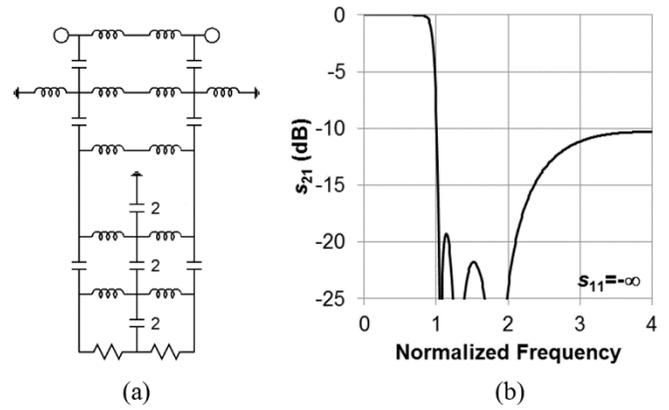

Fig. 9. a) Third-order nested reflectionless filter, with b) simulated performance. As with the original, all inductors, capacitors, and resistors have the same value, respectively, given in Section II, except those labeled "2", which have twice the nominal value.

### VI. Dual-Filter Sub-Networks

In Section II, the concept of dual circuits was discussed. Due to the symmetry of electric and magnetic fields in Maxwell's Equations, circuits and their duals have the same amplitude response, but with inverse normalized impedance characteristics. Since the normalized input impedance of a reflectionless filter is unity by design, one might expect it to be its own dual, however this is not the case. The dual first-order reflectionless low-pass filter is shown on the right in Fig. 10. It is in fact a circuit identity for the original filter topology shown on the left, having the same input impedance and frequency response, plotted in Fig. 2. (Interestingly, the dual reflectionless filter may also be obtained from the network in Fig. 6 by moving the terminations from the internal ports 3 and 4 to the external ports 1 and 2, and then redefining the "ground node" to one that is in common to ports 3 and 4.)

Besides academic curiosity, the significance of the dual reflectionless filter is revealed when one is embedded within the other as a sub-network. Depending on which of the two structures shown in Fig. 10 is embedded within the other, one finds that the resultant structures have either a series connection of parallel capacitor pairs, or a parallel connection of series inductor pairs. Either way, four identical elements are reduced to one. In this way, by alternating reflectionless filters and their duals in multiply-embedded structures, the final circuit topologies may be moderately simplified without changing the performance. A meaningful example is shown in Fig. 11, wherein the third-order nested reflectionless filter has been reduced in size by alternating the original reflectionless filter topology with its dual. These two circuits are identities for each other (as are the other six possible combinations one may obtain by nesting the structures shown in Fig. 10), having the same input impedance and frequency response as plotted in Fig. 9.

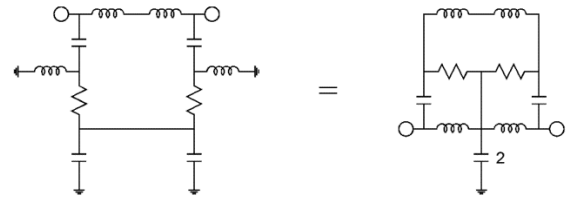

Fig. 10. Original reflectionless filter topology and its dual. The frequency response and input impedance are exactly the same as that shown Fig. 2.

### VII. Filters with Auxiliary Elements

It was shown in [1] that the symmetry and duality of the even- and odd-mode equivalent circuits, required for the "reflectionless" property, may still be satisfied with the addition of some auxiliary components, properly placed and valued complementary to one another, usually in the form of an inductor and capacitor pair. At the time of that publication, however, no constructive use for such additional components could be found, for by themselves, they did not seem to offer any improvement in the frequency response. With the advent of the sub-network technique described in this paper, however, useful combinations employing such additional components can be found. An example is shown in Fig. 12, where the original reflectionless filter topology with auxiliary components is embedded within the dual topology shown on the right side of Fig. 10. The value of the auxiliary components is a free parameter, but approximately 1/9th relative to nominal seems to be optimum in this case. The transition region is not quite as steep as the third-order filter, but the stop-band is much more extended.

### VIII. Inter-Cell Sub-Networks

Thus far the discussion has been limited to sub-networks embedded *within* a reflectionless filter cell. It is also possible to cross-connect the internal ports of two adjacent filter cells with a balanced inter-cell sub-network as shown in Fig. 13. As before, if the sub-network is itself matched, and is connected to the internal ports of the cascade as shown, then the composite filter structure will retain the reflectionless property. The introduction of this additional connection, however, introduces a closed loop in the signal flow path (see Fig. 13b) that can be used to alter the pass-band shape. If the sub-network transfer function is $Q(s)$, then the composite transfer function is given by

$$H'(s) = \frac{H^2(s)}{1 - P^2(s)Q(s)}. \qquad (8)$$

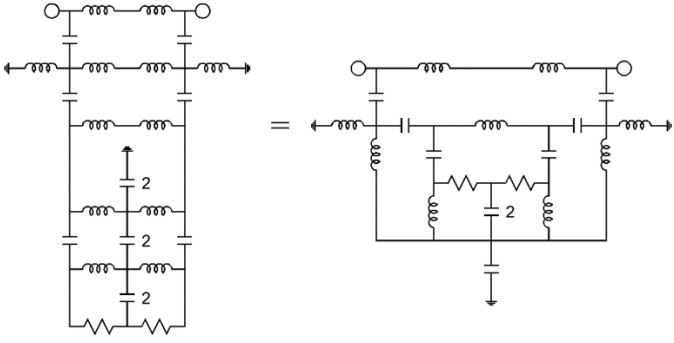

Fig. 11. Third-order nested reflectionless filter and its reduced equivalent circuit resulting from the substitution of the dual topology in the middle stage. Only two of eight possible circuits with identical performance are shown. The frequency response and input impedance are exactly the same as that shown in Fig. 9.

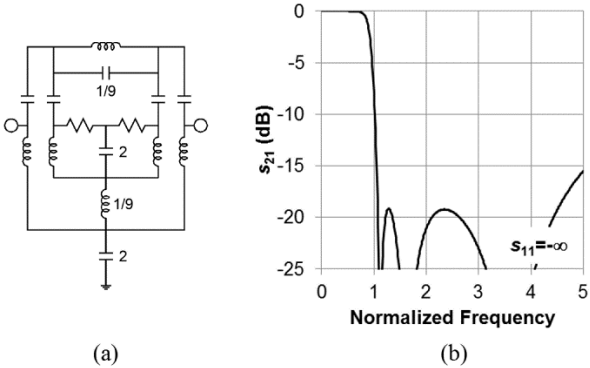

(a)      (b)

Fig. 12. a) Second-order reflectionless filter topology with auxiliary components (having values equal to 1/9th the nominal). To reduce the size, the original reflectionless filter has been embedded within its dual. b) Simulated performance.

For the 1:1 transformer, we have $Q(s)=1$, and the resulting frequency response is shown in Fig. 14. Note that the cutoff transition has been significantly sharpened while the overall stop-band rejection is commensurate with the expectation for a cascade of two reflectionless filters.

As in the previous section, the number of possible inter-cell networks, their combinations, and the manner in which groups of cells are interleaved and cross-connected is innumerable. One caveat, however, is that the cross-connection must remain totally differential and present a high-impedance to common-mode signals at the connection points. The transformer shown in the example of Fig. 13 is thus a necessity to avoid loading down the common-mode signals as would a direct-wire connection between the two-cells (resulting in a non-zero reflection coefficient for the overall structure).

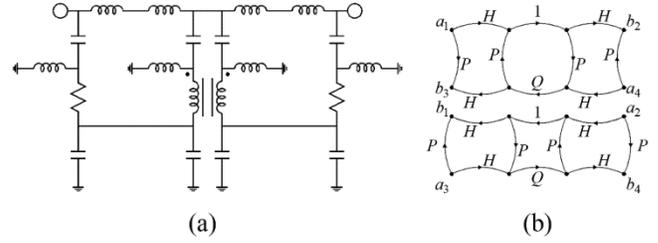

(a)      (b)

Fig. 13. a) Two cascaded reflectionless filter cells with a 1:1 transformer sub-network cross-connecting them. b) Signal-flow diagram.

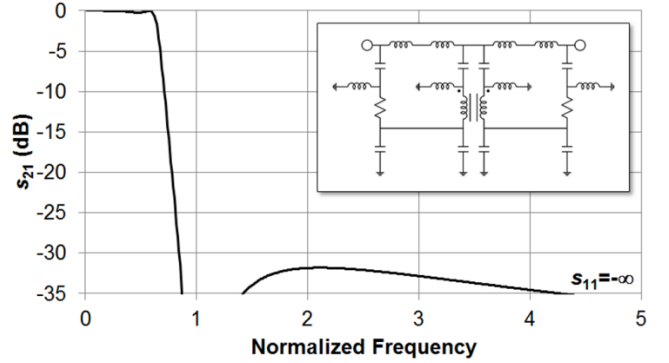

Fig. 14. Simulated response of the two-cell filter with transformer cross-connection.

## IX. Test Data

To verify the theory, as well as implement these filters in a manner that would be useful for targeted applications, several MMIC chips were fabricated based on the designs presented in this paper. The challenge of implementing these relatively complex filters in printed-circuit form is greatly alleviated by the fact that most if not all of the reactive elements are equivalued. This means that usually only a single spiral inductor needs to be optimized for a particular frequency, and then it can simply be replicated in the layout. The low quality factor (Q) of planar spiral inductors degrades the performance of the filter relative to the ideal-element model, but the convenience of being able to implement the filters in such a compact and mass-producible form is worthwhile in many applications.

The chips were fabricated using a passive-only GaAs MMIC process. A total of 23 designs was fabricated, comprising low-pass, high-pass, and band-pass designs, to meet the needs of a wide range of applications, as summarized in Table 1. The first 11 designs are of the basic low-pass type shown in Fig. 1d. One of these is illustrated in Fig. 15. Three more high-pass and a band-pass design were fabricated based on this topology after applying the necessary frequency domain transformation. Examples are shown in Fig. 16 and

TABLE I
MMIC REFLECTIONLESS FILTERS

| Part Number | Type | 3 dB Corner |
|---|---|---|
| MMRL500MD1 | low-pass | 300 MHz |
| MMRL625MD1 | low-pass | 385 MHz |
| *MMRL1G00D1 | low-pass | 630 MHz |
| MMRL1G25D1 | low-pass | 795 MHz |
| MMRL1G80D1 | low-pass | 1.20 GHz |
| MMRL2G50D1 | low-pass | 1.57 GHz |
| MMRL3G80D1 | low-pass | 2.51 GHz |
| MMRL5G00D1 | low-pass | 3.32 GHz |
| MMRL6G80D1 | low-pass | 4.25 GHz |
| MMRL16G0D1 | low-pass | 10.3 GHz |
| MMRL24G0D1 | low-pass | 13.7 GHz |
| *MMRH1G00D1 | high-pass | 1.65 GHz |
| MMRH1G25D1 | high-pass | 2.03 GHz |
| MMRH2G10D1 | high-pass | 3.27 GHz |
| *MMRB1G623G60D1 | band-pass | 2.3, 3.1 GHz |
| MMR3AD1 | 3rd-order low-pass | 8.2 GHz |
| MMR3BD1 | 3rd-order low-pass | 10.0 GHz |
| MMR3CD1 | 3rd-order low-pass | 11.3 GHz |
| MMR3DD1 | 3rd-order low-pass | 12.8 GHz |
| MMR3ED1 | 3rd-order low-pass | 13.6 GHz |
| MMR3FD1 | 3rd-order low-pass | 15.7 GHz |
| MMR3GD1 | 3rd-order low-pass | 16.9 GHz |
| *MMR3HD1 | 3rd-order low-pass | 21.1 GHz |

*All chips are 1.0 mm x 1.0 mm in size.*
*[*]Measurements are shown for the models identified with an asterisk.*

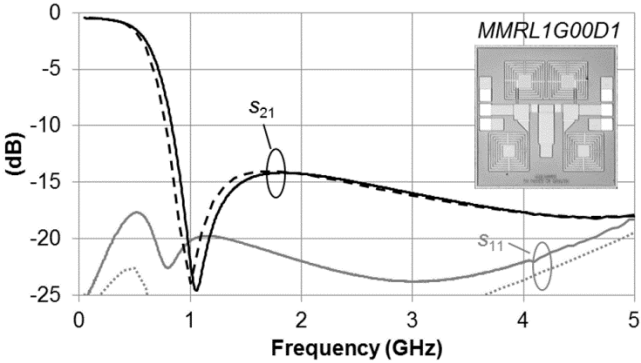

Fig. 15. EM-simulated (dashed/dotted) and measured (solid lines) performance of low-pass MMIC filter. Chip shown in photo inset is 1.0 mm x 1.0 mm x 0.1 mm in size.

Fig. 17.

It was evident from the measurements on these filters that measured results were tuned roughly 4% higher than expected from the 2.5D electromagnetic simulation of the complete structures using Sonnet Software. An investigation into the discrepancy, using the band-pass design in Fig. 17 as a reference, reveals that insufficient attention to the metal thickness is the probable cause. The initial simulations were performed using the "Normal" metal model (see dotted line in Fig. 17), wherein the traces are modeled electromagnetically as infinitely thin sheets, and a thickness is specified only for the purposes of calculating losses. The inductors in the filter structure, however, comprise metal spiral traces which are 4 μm thick while spaced only 6 μm apart. It is believed that the aspect ratio of the gaps is too small to use the thin-sheet model, and the "Thick Metal Model" provided by the software is required instead [8]. It was found in follow-up simulations using this model that slightly less than 1 μm thickness was sufficient to account for the frequency shift (see dashed line in

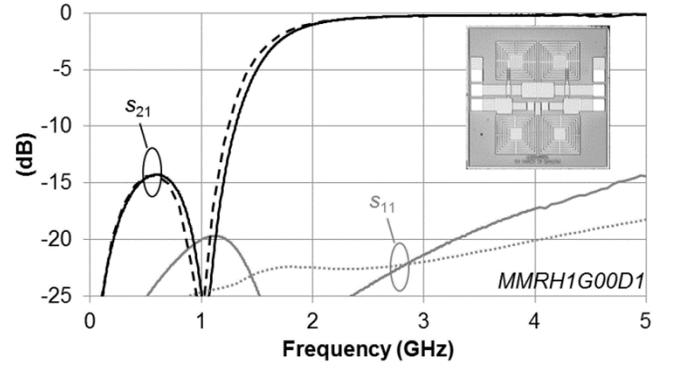

Fig. 16. EM-simulated (dashed/dotted) and measured (solid lines) performance of high-pass MMIC filter. Chip shown in photo inset is 1.0 mm x 1.0 mm x 0.1 mm in size.

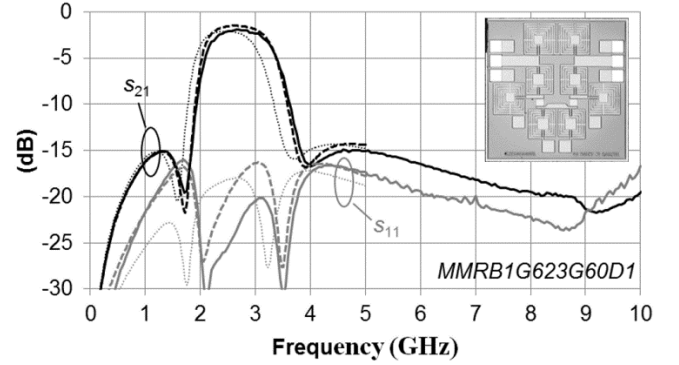

Fig. 17. EM-simulated with thin metal (dotted lines), with thick metal (dashed lines), and measured (solid lines) performance of band-pass MMIC filter. Chip shown in photo inset is 1.0 mm x 1.0 mm x 0.1 mm in size.

Fig. 17). The reason for the smaller than physical thickness required by the software can be explained by the expected shape of the metal trace cross-section, which in practice is usually more trapezoidal than rectangular.

The remaining eight filters fabricated in the initial run were of the third-order low-pass topology illustrated in Fig. 9. These were included at the request of a partner institution using them in active multiplier chains for the purpose of absorbing unwanted harmonics. A representative example of these filters is shown in Fig. 18, and further illustrates the high-frequency range over which these filters can be implemented despite the lumped-element realization.

The rapid generation of multiple layouts shown in Table I was greatly simplified by the duplication of element values inherent in the filter topologies, affording easy parameterization of the critical dimensions. Even the very complex third-order structures containing 12 inductors and 12 capacitors each required only two free parameters: the diameter of the inductors and the side-length of the capacitors. This allowed new tunings to be generated in seconds simply by entering a couple of numbers.

## X. Future Work

Some of the more sophisticated designs presented in this paper will be carried on future wafer runs. Layouts for several such filters have already been generated and modeled electromagnetically. The first is a reduced-element third-order filter based on nested dual topologies described in Section IV (shown on the right-hand side of Fig. 11). This should have the same passband shape as the eight third-order filters fabricated in Table 1 and shown by example in Fig. 18, but at much lower frequencies on the same size chip, thanks to the reduced number of elements of the equivalent topology. A simulation plot and layout of this chip is shown in Fig. 19.

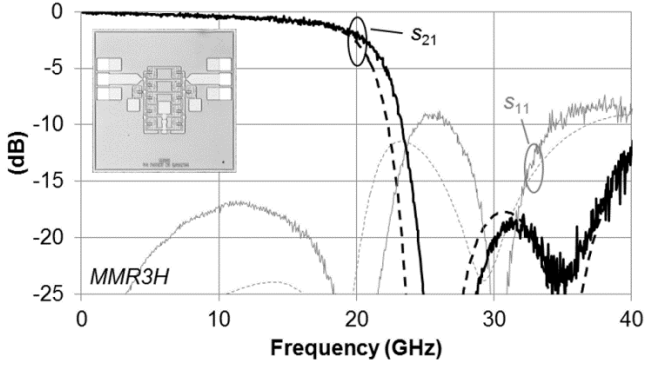

Fig. 18. EM-simulated (dashed lines) and measured (solid lines) performance of third-order low-pass MMIC filter. Chip shown in photo inset is 1.0 mm x 1.0 mm x 0.1 mm in size.

Another complex chip ready for fabrication is the second-order filter with auxiliary elements, the topology for which is shown in Fig. 12. A plot of the EM-simulated performance and layout is given in Fig. 20. The steepness of the transition region is only slightly less than the preceding filter, but with a much more extended stop-band.

## XI. Conclusion

This paper has presented a wide array of reflectionless filter structures, expanding on the basic design originally published [1]. For the first time, a transmission line equivalent structure has been presented, as has a powerful method for expanding the basic reflectionless filter structure to achieve a more sophisticated frequency response while retaining the reflectionless property. This method involves the introduction of matched embedded sub-networks with amplitude and phase that interfere with the primary leakage path in the stop-band of the embedding filter structure. The method may be generalized to arbitrarily complex networks involving hierarchically nested sub-networks, cascades, and cross-connections, with fairly simple rules to guarantee a reflectionless result. Several useful examples have been illustrated by simulation, and the basic concept confirmed by measurement of numerous low-pass, high-pass, and band-pass reflectionless filter structures implemented as a monolithic integrated circuits, and ranging from UHF to the lower mm-waves.

## Appendix

To prove the transmission line identity shown in Fig. 4, it is sufficient to show that the impedance (Z-) parameters of both networks are identical, given the relationships in (2). For the coupled-line section on the right of the figure, this is relatively straightforward – they are equivalent to the Z-parameters of a general four-port coupled transmission line with the fourth

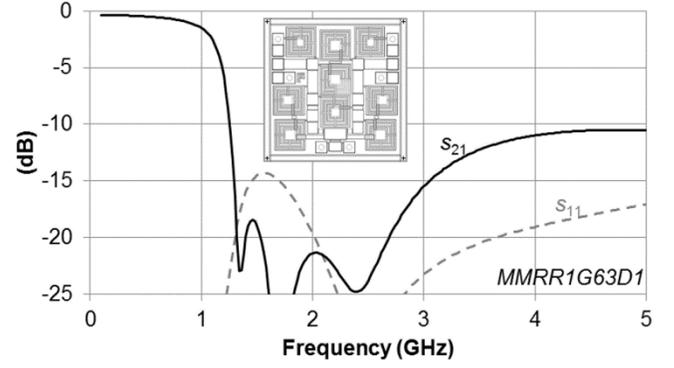

Fig. 19. EM-simulated performance of reduced third-order low-pass MMIC filter. Chip layout shown in inset is 1.0 mm x 1.0 mm x 0.1 mm in size.

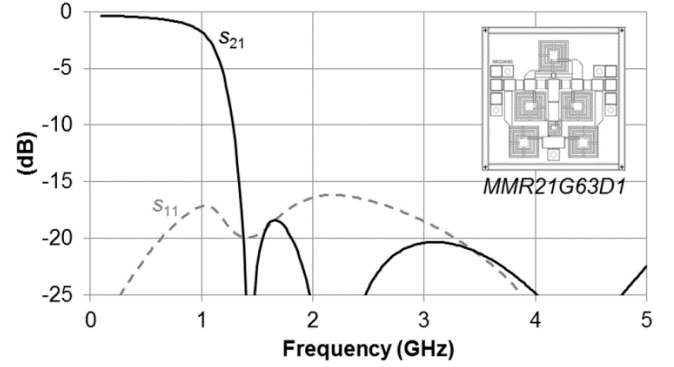

Fig. 20. EM-simulated performance of second-order low-pass MMIC filter with auxiliary components. Chip layout shown in inset is 1.0 mm x 1.0 mm x 0.1 mm in size.

port omitted (left open):

$$Z_{11} = Z_{22} = Z_{33} = \tfrac{1}{2}(Z_{even} + Z_{odd})\coth(\gamma l) \quad (9a)$$
$$Z_{12} = Z_{21} = \tfrac{1}{2}(Z_{even} + Z_{odd})\operatorname{csch}(\gamma l) \quad (9b)$$
$$Z_{13} = Z_{31} = \tfrac{1}{2}(Z_{even} - Z_{odd})\operatorname{csch}(\gamma l) \quad (9c)$$
$$Z_{23} = Z_{32} = \tfrac{1}{2}(Z_{even} - Z_{odd})\coth(\gamma l). \quad (9d)$$

Note that we have assumed TEM propagation – that is, the even- and odd-mode propagation velocities are the same ($\gamma_{even} = \gamma_{odd} = \gamma$). If we substitute (2) into these formulae,

$$Z_{11} = Z_{22} = Z_{33} = Z_1 \coth(\gamma l) \quad (10a)$$
$$Z_{12} = Z_{21} = Z_1 \operatorname{csch}(\gamma l) \quad (10b)$$
$$Z_{13} = Z_{31} = \tfrac{1}{n} Z_1 \operatorname{csch}(\gamma l) \quad (10c)$$
$$Z_{23} = Z_{32} = \tfrac{1}{n} Z_1 \coth(\gamma l). \quad (10d)$$

The task now is to derive the same expressions for the line and stub network on the left side of Fig. 4. If port 3 is left open, the remaining two-port network reduces to a simple transmission line with characteristic impedance $Z_1$. Therefore,

$$Z_{11} = Z_{22} = Z_1 \coth(\gamma l) \quad (11a)$$

$$Z_{12} = Z_{21} = Z_1 \operatorname{csch}(\gamma l) \quad (11b)$$

which matches (10b) and the first part of (10a). If ports 1 and 2 are left open, port 3 is looking into the series connection of two open-circuited stubs, modified by the transformer turns ratio,

$$Z_{33} = \tfrac{1}{n^2}\left(Z_1 \coth(\gamma l) + Z_2 \coth(\gamma l)\right)$$
$$= \frac{Z_1 + Z_2}{\frac{Z_2}{Z_1} + 1} \coth(\gamma l) = Z_1 \coth(\gamma l) \quad (12)$$

which verifies the remainder of (10a). For the remaining parameters, we consider the voltage that is developed at ports 1 and 2 when a current, $I_3$, is injected into port 3. We observe that the transformer reduces this current to $I_3/n$, and further that the open stub with characteristic impedance $Z_2$ is in series with this current source and can therefore be neglected. We are thus left once again with a two-port transmission line having characteristic impedance $Z_1$, but with the current source at port 2 reduced by a factor $1/n$,

$$Z_{13} = Z_{31} = \tfrac{1}{n} Z_1 \operatorname{csch}(\lambda l) \quad (13a)$$
$$Z_{23} = Z_{32} = \tfrac{1}{n} Z_1 \coth(\lambda l). \quad (13b)$$

This confirms (10c) and (10d), thus proving the identity in Fig. 4.

## Acknowledgment

This research was carried out at the National Radio Astronomy Observatory (NRAO), Charlottesville, VA, a facility of the National Science Foundation (NSF) operated under cooperative agreement by Associated Universities Inc.

**Matthew A. Morgan** (M'99) received the B.S.E.E. degree from the University of Virginia, Charlottesville, in 1999, and the M.S. and Ph.D. degrees in electrical engineering from the California Institute of Technology, Pasadena, in 2001 and 2003, respectively.

From 1999 to 2003, he was an Affiliate of the Jet propulsion Laboratory (JPL), Pasadena, CA, where he developed monolithic millimeter-wave integrated circuits (MMICs) and multi-chip modules (MCMs) for atmospheric radiometers and spacecraft telecommunication systems.

Dr. Morgan is currently a Scientist / Research Engineer with the National Radio Astronomy Observatory (NRAO), Charlottesville, VA, where he is involved in the design and development of low-noise receivers, components, and novel concepts for radio astronomy instrumentation in the cm-wave, mm-wave, and submm-wave frequency ranges. He was Project Engineer for the K-Band Focal Plane Array on the Green Bank Telescope, and is currently the head of CDL's Integrated Receiver Development Program.

**Tod. A. Boyd** was born in Steubenville, Ohio, in 1962. He received the A.S.E.E. degree from the Electronic Technology Institute, Cleveland, Ohio, in 1983. From 1983 to 1985, he was with Hostel Electronics in Steubenville, Ohio. In 1985, he joined Northrop Corporation's Electronic Countermeasures Division, in Buffalo Grove, Ill., specialized in supporting the B-1B Lancer (secret clearance). In 1990, he joined Interferometrics, Inc., Vienna, Va., where he constructed VLBA tape recorders for the international Radio Astronomy community.

Since 1996, he has been with the National Radio Astronomy Observatory's Central Development Lab, Charlottesville, VA, where he initially assisted with the construction of cooled InP HFET amplifiers for the NASA's Wilkinson Microwave Anisotropy Probe (WMAP) mission. Presently as a Technical Specialist I he provides technical support for the advanced receiver R&D initiatives. His responsibilities also include constructing low noise amplifiers for the Enhanced VLA and the Atacama Large Millimeter/sub-millimeter Array (ALMA) projects.